\input{aipcheck}

\documentclass[
    ,final            % use final for the camera ready runs
  ]
  {aipproc}

\layoutstyle{8x11single}

\def\ltap{\ \raise.3ex\hbox{$<$\kern-.75em\lower1ex\hbox{$\sim$}}\ }
\def\gtap{\ \raise.3ex\hbox{$>$\kern-.75em\lower1ex\hbox{$\sim$}}\ }

\begin{document}

\title{Toward Construction of the Unified Lepton-Nucleus Interaction Model from a Few Hundred MeV to GeV Region}

% 25.30.Pt : Neutrino-induced reactions (on nuclei)
%13.15.+g : Neutrino interactions
%12.15.Ji : Applications of electroweak models to specific processes 


\author{S. X. Nakamura}{
  address={Yukawa Institute for Theoretical Physics, Kyoto University, Kyoto 606-8542, Japan}
}

\author{Y. Hayato}{
  address={Kamioka Observatory, Institute for Cosmic Ray Research, University of Tokyo, Kamioka, Japan}
}

\author{M. Hirai}{
  address={Department of Physics, Tokyo University of Science,  Noda 278-8510, Japan}
}

\author{H. Kamano}{
  address={Research Center for Nuclear Physics, Osaka University, Ibaraki 567-0047, Japan}
}

\author{S. Kumano}{
  address={KEK Theory Center, Institute of Particle and Nuclear Studies, KEK, Tsukuba 305-0801, Japan}
  ,altaddress={J-PARC Branch, KEK Theory Center, Institute of Particle and Nuclear Studies, KEK, Tokai 319-1106, Japan}
}

\author{M. Sakuda}{
  address={Department of Physics, Okayama University, Okayama 700-8530, Japan}
}

\author{K. Saito}{
  address={Department of Physics, Tokyo University of Science,  Noda 278-8510, Japan}
}

\author{T. Sato}{
  address={Department of Physics, Osaka University, Toyonaka, Osaka 560-0043, Japan}
  ,altaddress={J-PARC Branch, KEK Theory Center, Institute of Particle and Nuclear Studies, KEK, Tokai 319-1106, Japan}
}

\begin{abstract}
Next generation neutrino oscillation experiments will need a
 quantitative understanding of neutrino-nucleus interaction far better
 than ever. 
Kinematics covered by the relevant neutrino-nucleus interaction
 spans wide region, from the quasi-elastic, through the resonance
 region, to the deeply inelastic scattering region.
The neutrino-nucleus interaction in each region has quite different
characteristics.
Obviously, it is essential to combine different expertise to construct a
 unified model that covers all the kinematical region of
the neutrino-nucleus interaction.
Recently, several experimentalists and theorists got together to form a
 collaboration to tackle this problem. 
In this contribution, we report the collaboration's recent activity and a goal in near
 future. 
\end{abstract}

\maketitle

%%%%%%%%%%%%%%%%%%%%%%%%%%%%%%%%%%%%%%%%%%%%
%% MAINMATTER
%%%%%%%%%%%%%%%%%%%%%%%%%%%%%%%%%%%%%%%%%%%%

\section{Introduction}

\begin{figure}
\includegraphics[width=7cm, bb=0 0 396 409]{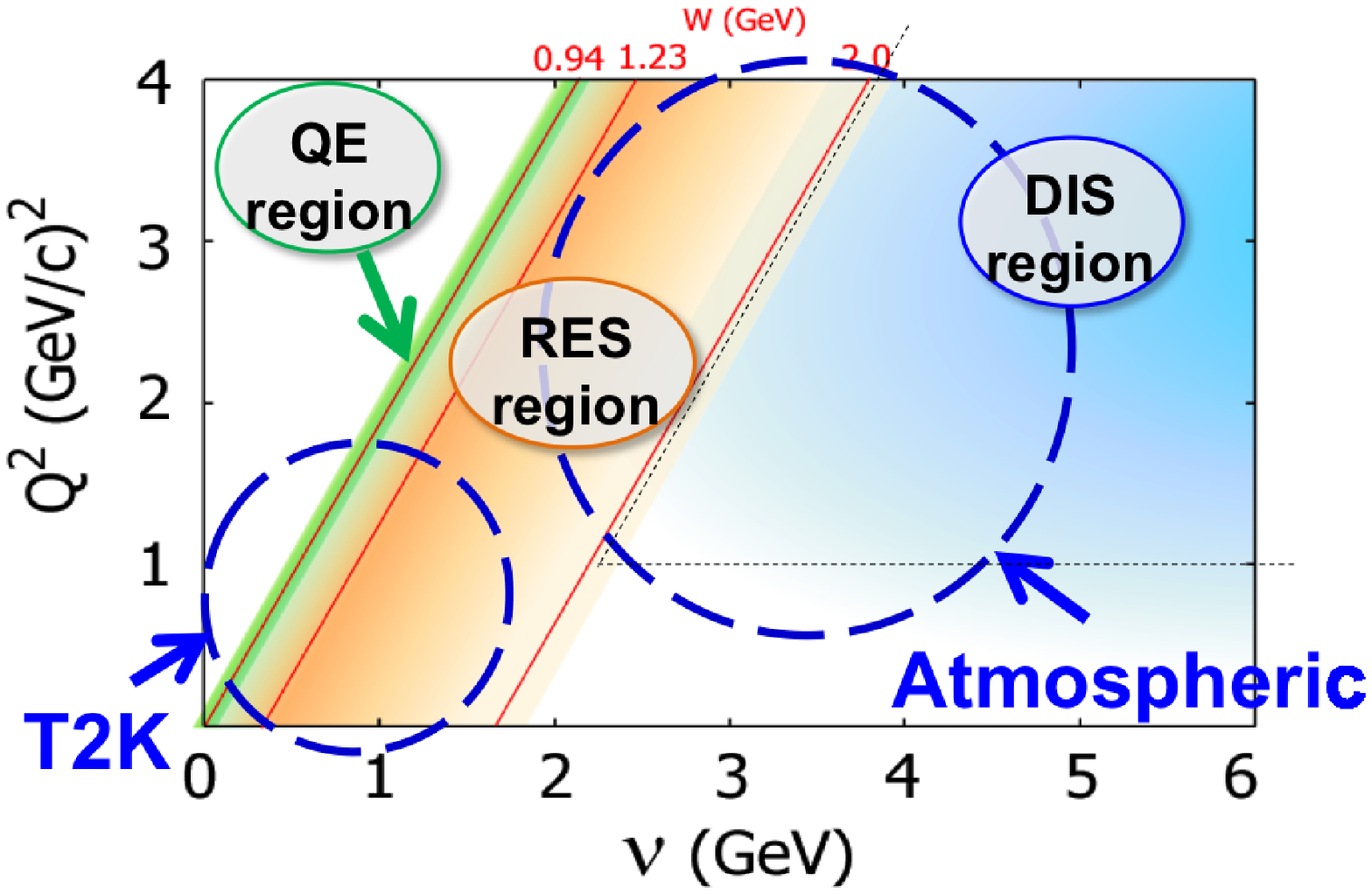}
\hspace{20mm}
\includegraphics[width=6.5cm, bb=0 0 396 409]{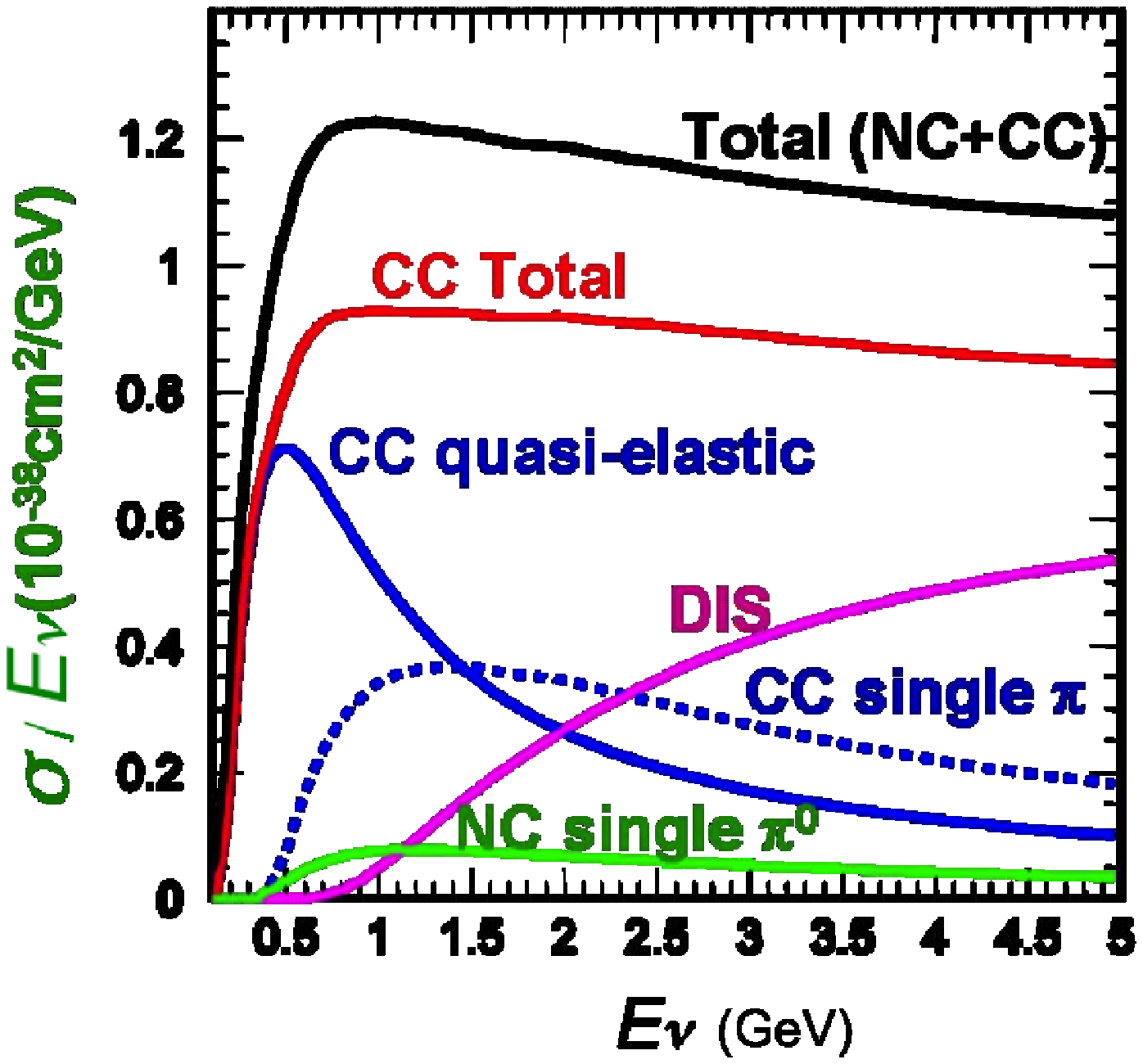}
  \caption{\label{fig1} 
(Left) Kinematical region of neutrino-nucleus interaction 
relevant to the next-generation neutrino oscillation experiments.
The neutrino energy and squared four-momentum transfer are denoted by $\nu$
 and $Q^2$, respectively. 
(Right) Contribution of each reaction mechanism to the neutrino-nucleus
 interaction. 
The symbol $\sigma$ denotes the cross section.
}
\end{figure}

The breakthrough measurement of non-zero $\theta_{13}$ is changing
the neutrino oscillation experiments into a new phase.
The next target will be the leptonic CP violation, and the mass
hierarchy of the neutrino. 
For aiming at this target, we need not only better controlled, higher statistics experiments 
but also a quantitative understanding of
neutrino-nucleus interaction at the level of 5\% or better.
The relevant neutrino-nucleus interaction
covers a wide kinematical region, as described in Fig.~\ref{fig1} (left).
From the low to high energy side, characteristics of the
neutrino-nucleus interaction changes:
from the quasi-elastic (QE), through the resonance (RES)
region, to the deeply inelastic scattering (DIS) region.
Contribution from each reaction mechanism to the neutrino-nucleus
interaction is displayed in Fig.~\ref{fig1} (right).
There, CC (NC) stands for the charged-current (neutral-current).
The single $\pi$ processes are associated with the resonance excitation.
A long baseline accelerator neutrino experiment like T2K utilizes the
neutrino-nucleus interaction at the left-bottom corner of
Fig.~\ref{fig1} (left) where QE and the single pion production due to the
$\Delta$ resonance excitation are main mechanism.
Meanwhile, an atmospheric neutrino experiment 
expects to find effects of CP-violation/mass hierarchy 
in the higher resonance and DIS regions, as indicated in the figure.
Obviously, it is essential to combine different expertise to construct a
 unified model covering all the kinematical region for the 
 neutrino-nucleus interaction.

Recently, several experimentalists and theorists got together to form a
 collaboration at J-PARC branch of KEK Theory Center to tackle this problem~\cite{jparc}. 
Our strategy is to develop and combine baseline models for QE, RES and DIS.
In this contribution, we report the current status of the baseline models
 for QE, RES and DIS, and discuss a future perspective.

\section{Quasi-Elastic region}

For the neutrino energy between 0.1 GeV and 1 GeV, the QE process dominates the 
neutrino-nucleus interaction, as shown in Fig.~\ref{fig1} (right).
At T2K energy, this QE process is the most dominant and important.
Since the neutrino oscillations depend on the energy of the neutrino,
the neutrino oscillation experiments 
must reconstruct the neutrino energy by selecting events from the CC QE processes.

At the summary session of the NuInt12 Workshop, we generally agreed that the spectral function of nuclei
should be used to calculate  QE process and pion production in the neutrino-nucleus interactions, rather
than a simple uniform nucleon momentum distribution based on the Fermi-Gas model, in order to 
achieve better accuracy of the calculations.  
The nuclear spectral functions $P(p,E)$ is the probability of 
removing a nucleon of momentum ($p$) from ground state
of the nucleus leaving the residual nucleus with excitation energy ($E$).
The nuclear spectral function naturally enables us not only to calculate both CC and 
NC QE cross section,
but also to estimate a spectroscopic factor ($S_p$) that is
the probability of the residual nucleus in a specific shell state.
Basic calculations with the spectral function approach
are fully described in Ref.\cite{QE}. 
Comparisons of electron-nucleus scattering data with the 
spectral function approach and with the Fermi-Gas model are also shown in Ref.\cite{QE}.  
This approach is also useful to calculate 
CC and NC QE processes followed by a $\gamma$ emission~\cite{QEgamma}
for which the branching fraction is significant ($\sim 40$\%).

A remaining issue is to better understand effects of 
2p-2h (2-particle-2-hole) configuration and/or MEC (meson-exchange
current) in the QE process.
This subject has been raised by several recent experiments on
the QE cross section measurement~\cite{qe_exp}.
The data seem to indicate a considerable contribution from 2p-2h and/or
MEC effects, and have motivated active theoretical studies~\cite{qe_th}.

\section{Resonance region}

The neutrino-nucleon interaction in the resonance region is a
multi-channel reaction where not only single pion production but also 
two-pion production has a comparable contribution above the
$\Delta(1232)$.
$\eta$ and kaon productions can also happen with a smaller probability.
In order to deal with this kind of multi-channel reaction, 
an ideal approach is to develop a unitary coupled-channels model.
In the following subsections, we discuss the unitary dynamical
coupled-channels (DCC) model, and its extension to forward
neutrino-induced meson productions.\\

\noindent {\bf Dynamical coupled-channels model}\\
In the DCC model~\cite{msl,aip11,knls12}, 
we consider 8 channels: 
$\gamma N, \pi N, \eta N, \pi\Delta, \rho N, \sigma N, K\Lambda, K\Sigma$.
We solve a coupled-channel Lippmann-Schwinger equation 
that contains meson-exchange potentials and bare $N^*$ excitation mechanisms,
thereby obtaining unitary reaction amplitudes.
We analyzed 
$\pi(\gamma) N\to \pi N, \eta N, K\Lambda, K\Sigma$ reactions data 
simultaneously up to $W = 2.1$~GeV ($W$ : total energy).
The analysis includes fitting about 20,000 data points.
%
%As an example for showing the quality of the fit,
%we show in Fig.~\ref{fig2}
%the single pion photoproduction observables from the DCC model
%compared with data.
%
%\begin{figure}
%  \includegraphics[width=10cm]{g1p}
%  \caption{\label{fig2} Comparison of the DCC model and data for
%single pion photoproduction~\cite{knls12}.
%Total cross sections ($\sigma$), unpolarized
% differential cross sections ($d\sigma/d\Omega$) and photon asymmetry
% ($\Sigma$) are shown.
%The total energy is denoted by $W$, and the scattering angle of the
% pion by $\theta$.
%}
%\end{figure}
%
%As shown in the figure, 
%our DCC model gives a reasonable description of meson production
%data in the resonance region.
%Thus the DCC model provides a good basis with which we proceed to the neutrino reactions.\\
%%
The DCC model gives a reasonable description of meson production
data in the resonance region, thus providing 
a good basis with which we proceed to the neutrino reactions.\\

\begin{figure}
  \includegraphics[height=.25\textheight]{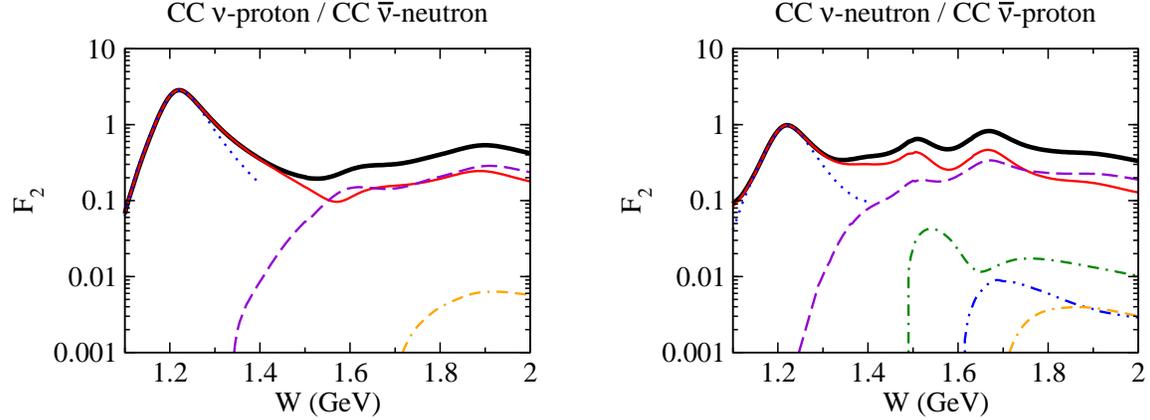}
  \caption{\label{fig3} The structure function $F_2 (Q^2=0)$ for the
 neutrino-induced meson productions from the DCC model. 
The solid (red), dashed (purple), dash-dotted (green), two-dotted
 dash (blue), and two-dash dotted (orange) curves are for the $\pi N$,
$\pi\pi N$, $\eta N$, $K\Lambda$ and $K\Sigma$ reactions, respectively.
The sum of them is given by the thick solid (black) curve. 
%The Sato-Lee model is shown by the dotted (blue) curve.
}
\end{figure}

\noindent{\bf Forward neutrino-induced meson productions based on PCAC}\\
In the forward limit, the matrix element for
a neutrino-induced meson production off the nucleon is given by the
divergence of the axial current, accompanied by some kinematical
factor. 
With the Partially Conserved Axial Current (PCAC) hypothesis, we can
relate the matrix element with those of the pion-induced meson
production.
Thus with $\pi N\to X$ ($X=\pi N, \pi\pi N, \eta N, K\Lambda, K\Sigma$)
amplitude, we can calculate the forward
neutrino-induced meson productions. 
Recently, We have done such an application with the DCC model for $\pi N\to X$~\cite{pcac}.
We present the result in Fig.~\ref{fig3} where the structure function 
$F_2 (Q^2=0)$ is shown as a function of the total energy $W$ of the
hadronic final state.
While $\pi N$ production is the dominant process up to $W=1.5$ GeV, 
above that energy, 
the $\pi \pi N$ production becomes comparable to $\pi N$, showing the
importance of the $\pi\pi N$ channel in the resonance region above $\Delta(1232)$.
Other meson productions, $\eta N$, $K\Lambda$, and $K\Sigma$ reactions
have much smaller contribution.
We remark that this is the first prediction of 
the neutrino-induced $\pi\pi N$, $\eta N$, $KY$ production rates based
on a model that has been extensively tested by data. \\

\noindent{\bf Future plan}\\
The current model for the neutrino reaction 
is limited to only the forward kinematics.
In near future, we extend the model so that we
can analyze the neutrino reaction of all kinematics in the resonance region.
For that, we need to develop a dynamical model for both vector and axial currents.
Regarding the vector current, we already have some part by having analyzed data for meson
photoproductions on the proton. 
We will also analyze the neutron target data, so that we can decompose the
vector current into isovector and isoscalar pieces; 
this decomposition is necessary for calculating the neutrino processes.
The $Q^2$-dependence of the vector current couplings will be fixed 
by analyzing pion electroproduction data.
For the axial current, we develop a model consisting of
non-resonant mechanisms and also bare
$N^*$ excitation mechanisms. 
The axial current couplings of $N$-$N^*$ transition can be determined
from the corresponding $\pi N N^*$ coupling using the PCAC relation.
The $Q^2$-dependence of the couplings can be fixed using available data
of the neutrino-induced single pion production.
In this way, we obtain all ingredients to calculate the neutrino-induced
meson productions for finite $Q^2$ in the resonance region.

%%%%%%%%%%%%%%%%%%%%%%%%%%%%%%%%%%%%%%%%%%%%%%%%%%%%%%%%%%%%%%%%%%%%%%%%%%%%%%%%
\section{Deep inelastic scattering region}

The deep inelastic region of neutrino-nucleon interactions 
corresponds to the kinematical conditions of $Q^2 \gtap 1$ GeV$^2$ and
$W^2 \gtap 4$ GeV$^2$, and the cross sections or structure functions 
are described by the parton model.
The neutrino-nucleon cross section via charged current is
expressed by the three structure functions $F_1$, $F_2$, and $F_3$:
\begin{equation}
\left( {\frac{{d\sigma }}{{dxdy}}} \right)_{CC}^{\nu ,\bar \nu } 
= \frac{{G_F^2 M_N E}}{\pi{(1 + Q^2 /M_W^2 )^2 }}
\left[ {F_1^{cc} xy^2  + F_2^{cc}
\left( {1 - y - \frac{{M_N xy}}{{2E}}} \right) 
\pm F_3^{cc} xy\left( {1 - \frac{y}{2}} \right)} \right] ,
\end{equation}
where $\pm$ indicates $+$ ($-$) for $\nu$ ($\bar \nu$),
$x$ is the Bjorken scaling variable $x=Q^2/(2 p\cdot q)$
with the nucleon momentum $p$, the $W$-boson momentum $q$,
and $Q^2 = -q^2$, 
$M_N$ and $M_W$ are the nucleon and $W$-boson masses,
$y$ is defined by $y=p\cdot q/(p\cdot k)$
where $k$ is the neutrino momentum,
and $E$ is the neutrino-beam energy.
In the leading twist and the leading order (LO) of the running
coupling constant $\alpha_s$,
the structure functions are expressed in terms of 
parton distribution functions (PDFs):
\begin{eqnarray}
%            2xF_1  & = F_2, \ \ \ \ \  &
%           \ \ \   & \ \ \
% \nonumber   \\
2xF_1  = F_2, \ \ \ \ \
      F_2^{\nu p}   = 2x(d + s + \bar u + \bar c), \ \ \ \ \  
 F_2^{\bar \nu p}   = 2x(u + c + \bar d + \bar s),     
\nonumber   \\ 
      xF_3^{\nu p}  = 2x(d + s - \bar u - \bar c), \ \ \ \ \  
 xF_3^{\bar \nu p}  = 2x(u + c - \bar d - \bar s) .   
\label{eqn:f123}
\end{eqnarray}
The next-to-leading order (NLO) expressions are calculated by
the convolution integral of the above PDFs with NLO coefficient functions.
%%%%%
For describing neutrino-nucleus cross sections,
the nucleonic structure functions $F_{1,2,3}$ should be replaced
by the corresponding nuclear ones $F_{1,2,3}^{A}$, which
include nuclear corrections.
The nuclear structure functions are expressed by
nuclear PDFs (NPDFs) $f_i^A (x)$ for the parton type $i$
in the same way as Eq. (\ref{eqn:f123}).
They contain nuclear correction factors $w_i (x,A,Z)$,
which indicate deviations from a simple addition of
proton ($p$) and neutron ($n$) contributions:
\begin{equation}
f_i^A (x,Q_0^2) = w_i (x,A,Z) \, 
    \frac{1}{A} \left[ Z\,f_i^{p} (x,Q_0^2) 
                + (A - Z) f_i^{n} (x,Q_0^2) \right] ,
\end{equation}
where $Z$ is the atomic number of a nucleus, $A$ is the mass number,
and $Q_0^2$ is the initial $Q^2$ scale which is taken
as $Q_0^2=1$ GeV$^2$.
The weight functions $w_i (x,A,Z)$ are determined by
a global analysis of experimental data on 
lepton-nucleus deep inelastic scattering (DIS) and
Drell-Yan processes with nuclear targets \cite{npdfs}.

Neutrino DIS data can be included
in the global analysis for extracting the NPDFs. However,
one should be careful that nuclear modification data such
as $F_2^A /F_2^D$, where $D$ indicates the deuteron,
are not available in the neutrino scattering. 
Recently, there are some discussions on possible differences
between nuclear modifications of charged-lepton and neutrino
reactions. In the MINER$\nu$A project, structure functions
of light nuclei will be measured, so that such an issue will be
investigated independently.

In Fig.~1 (left), we notice a region which does not belong to
either the resonance region or the DIS one at 
$Q \le 1$ GeV$^2$ and $W^2 \ge 4$ GeV$^2$.
In order to provide a complete interaction model for
various neutrino reactions, we need to describe such a region.
At $Q^2 \to 0$, there is a guiding principle like 
PCAC
to connect a structure function to the pion scattering 
cross section. Together with the update of the NPDFs
including new data, an appropriate description of this
regions should be developed. These studies are currently in progress.

%%%%%%%%%%%%%%%%%%%%%%%%%%%%%%%%%%%%%%%%%%%%%%%%%%%%%%%%%%%%%%%%%%%%%%%%%%%%%%%%

%%%%%%%%%%%%%%%%%%%%%%%%%%%%%%%%%%%%%%%%%%%%%%%%
%% BACKMATTER
%%%%%%%%%%%%%%%%%%%%%%%%%%%%%%%%%%%%%%%%%%%%%%%%

\begin{theacknowledgments}
The authors thank the J-PARC branch of the KEK theory
center for supporting the collaboration's activity.
\end{theacknowledgments}



\bibliographystyle{aipproc}   % if natbib is available

%\bibliography{sample}

%  \typeout{******************************************}
%  \typeout{** Please run "bibtex \jobname" to optain}
%  \typeout{** the bibliography and then re-run LaTeX}
%  \typeout{** twice to fix the references!}
%  \typeout{******************************************}

%%%%%%%%%%%%%%%%%%%%%%%%%%%%%%%%%%%%%%%%%%%
%% The following lines show an example how to produce a bibliography
%% without the help of the BibTeX program. This could be used instead
%% of the above.
%%%%%%%%%%%%%%%%%%%%%%%%%%%%%%%%%%%%%%%%%%%

\end{document}